\documentclass[%
 aip,
 rsi,%
 amsmath,amssymb,
 reprint,%
groupedaddress
]{revtex4-1}
\usepackage{natbib}
\usepackage{color}

\usepackage{graphicx}
\usepackage{dcolumn}
\usepackage{bm}

\usepackage{siunitx}

\begin{document}


\title{A short response-time atomic source for trapped ion experiments}

\author{T. G. Ballance}
\author{J. F. Goodwin}
 \email{joseph.goodwin@physics.ox.ac.uk}
\author{B. Nichol}
\author{L. J. Stephenson}
\author{C. J. Ballance}
\author{D. M. Lucas}
\affiliation{Clarendon Laboratory, Parks Rd, Oxford, OX1 3PU, UK}

\date{\today}

\begin{abstract}
Ion traps are often loaded from atomic beams produced by resistively heated ovens. We demonstrate an atomic oven which has been designed for fast control of the atomic flux density and reproducible construction. We study the limiting time constants of the system and, in tests with $^{40}\textrm{Ca}$, show we can reach the desired level of flux in $\SI{12}{\second}$, with no overshoot. Our results indicate that it may be possible to achieve an even faster response by applying an appropriate one-off heat treatment to the oven before it is used.
\end{abstract}

\maketitle

\
\section{Introduction}
Quadrupole ion traps find use in many areas of physics including the study of fundamental constants~\cite{Huntemann2014, Godun2014}, building accurate clocks~\cite{Ludlow2015}, and as platforms for quantum computation~\cite{Cirac1995}.

In these experiments, loading of the trap is usually accomplished by producing ions inside the trapping region from an atomic vapor, either by electron impact ionization or resonance-enhanced two-photon ionization\cite{Kjaergaard2000}, the latter providing a means for isotope-selective loading. Owing to the need for ultra-high vacuum conditions during operation of the ion trap, the vapor is usually provided by a collimated atomic beam from an oven which can be resistively heated when loading is desired.

An alternative method for producing an atomic vapor for loading is laser ablation of a coated target~\cite{Knight1981}. Typically, a pulsed laser will be focused onto the a target within line of sight of the trap. A single pulse with sufficient energy will ablate the surface of the target producing an atomic vapor. Once a vapor is present within the trapping region, atoms are ionized using either electron impact or resonant photoionization. This method has the advantage that the atomic flux can be created `on demand' on the millisecond timescale.

Recent experiments have also demonstrated loading via ionization of a cold atomic beam emitted from a magneto-optical-trap (MOT) adjacent to the ion trap~\cite{Sage2012,Bruzewicz2016}. Because the atoms are pre-cooled before ionization, efficient loading of shallow surface traps can be achieved with a far lower atomic flux than is produced by a conventional thermal oven.

As the applications of ion traps become increasingly diverse and the complexity of experiments rises, it is important to identify elements of the apparatus where well-engineered solutions can help produce reliable, stable and consistent performance across multiple units with minimal calibration.
In order to load ion traps with complex multi-species ion chains it is important that the loading process meets certain requirements:
\begin{enumerate}
\item{
the system is capable of loading single ions of a chosen isotope with a high reliability,
}
\item{
the atomic flux incident on the trap surfaces is minimized, to reduce contamination which may generate stray fields or increased motional heating,\cite{Sage2012}
}
\item{
the loading mechanism is simple to reproduce consistently across many systems,
}
\item{
the time spent loading must not be a significant overhead in the operation of the system.
}
\end{enumerate}

The final requirement is easily met in most current ion trap experiments owing to large storage times and relatively low ion numbers. However, when multi-species chains of several ions are desired\cite{Nigmatullin2016}, loading time can easily become an experimental bottleneck.
For example, in a surface trap containing a string of 5-10 ions, a loss event will typically occur every few minutes.
Typically, the resistively heated atomic ovens employed in our lab have required \numrange{2}{5} minutes to heat up from room temperature, which would significantly impact the rate at which experimental data could be collected. If the oven is left on, the rise in background pressure causes an undesirable decrease in trap lifetime .

Although attractive for the simplicity of its in-vacuum components, extremely fast loading time, and good compatibility with cryogenic vacuum systems, laser ablation may not be suitable for all systems. The ablation plume produced by the pulse contains mainly neutral atoms, in combination with a small number of atomic and molecular ions of a variety of species and isotopes\cite{Zimmermann2012}. Combined with the high kinetic energy of the ablation products, this limits the loading efficiency and will lead to an increased depletion rate of the ablation target\cite{Zimmermann2012, Guggemos2015}. Combining ablation with resonant photoionization increases efficiency and selectivity, but may not completely avoid trapping of other charged ablation products.

MOT-based loading schemes are appealing due to their greatly reduced flux per loaded ion, but require more laser power and strong but well-shielded magnetic fields. These elements introduce additional experimental complexity and physical size, which may be undesirable when attempting to produce compact systems for use in larger-scale quantum computing devices.

In this paper, we present an atomic source which can be resistively heated from room temperature in 12 seconds to produce an atomic beam with flux density suitable for loading an ion trap with resonant photoionization. We characterize the atomic flux by measuring resonance fluorescence from spectroscopy on the $4\textrm{s}^2{^1\textrm{S}_0}\to4\textrm{s}4\textrm{p}^1\textrm{P}_1$ transition in atomic calcium near \SI{423}{nm}. Full schematics for the oven and the accompanying electronic controller are provided online.\cite{GrimRepo}

\section{Construction}

The atomic source presented here consists of a resistively heated stainless steel tube (outer diameter \SI{1}{mm}, wall thickness \SI{50}{\um}, length approximately \SI{15}{mm}). The tube is held at either end by two flexure clamps made from stainless steel, which also form the electrical contacts. The central section of tube between the two clamps is crimped closed at two points, and the enclosed volume filled with calcium metal of natural isotopic abundance before the second crimp is made. Near the centre of the crimped section of tube is a \SI{0.5}{mm} diameter aperture which allows the calcium vapor to escape when the tube is heated.
The flexure clamps are mounted on an electrically and thermally isolating structure made from Macor, providing a stable and reproducible construction. This part contains a \SI{2}{mm} diameter hole which is used for collimation of the atomic beam. The distance between the oven aperture and the exit of the collimation aperture is \SI{13.5}{mm}, producing an atomic beam with a half-angle divergence of approximately 5 degrees.

\begin{figure}
\includegraphics[trim={10cm 10cm 12cm 6cm},clip,width=\columnwidth]{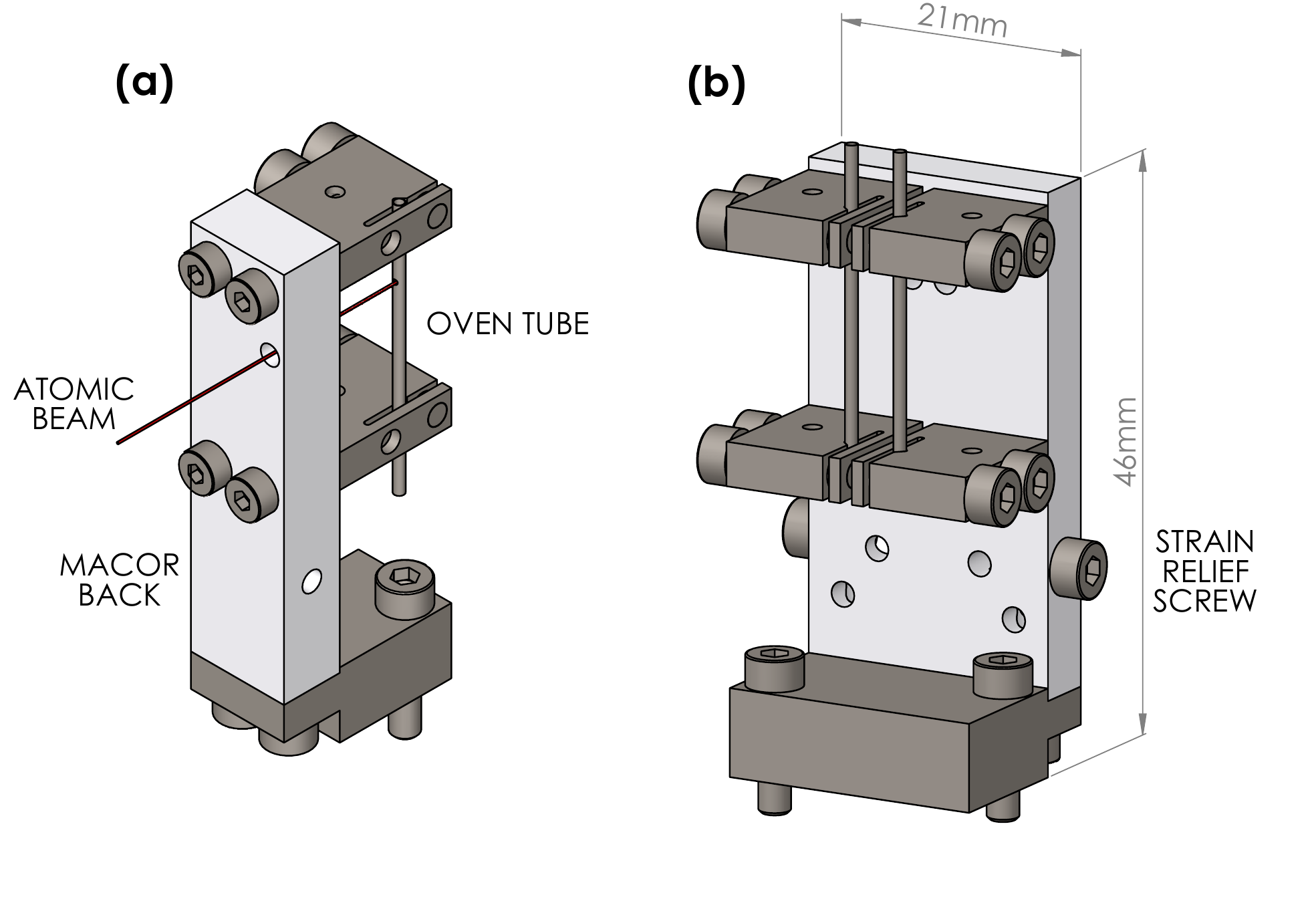}
\caption{Overview of the assembled oven mounts. (a) Single oven design used in these results, (b) Dual oven design suitable for multi-species experiments. Thermocouples are not shown for clarity, but are spot-welded to the oven tube and supported by the strain relief screws on the edges of the Macor block.}
\label{fig:oven_assy}
\end{figure}

\begin{figure}
\includegraphics[trim={1cm 19cm 5cm 3.5cm},clip,width=\columnwidth]{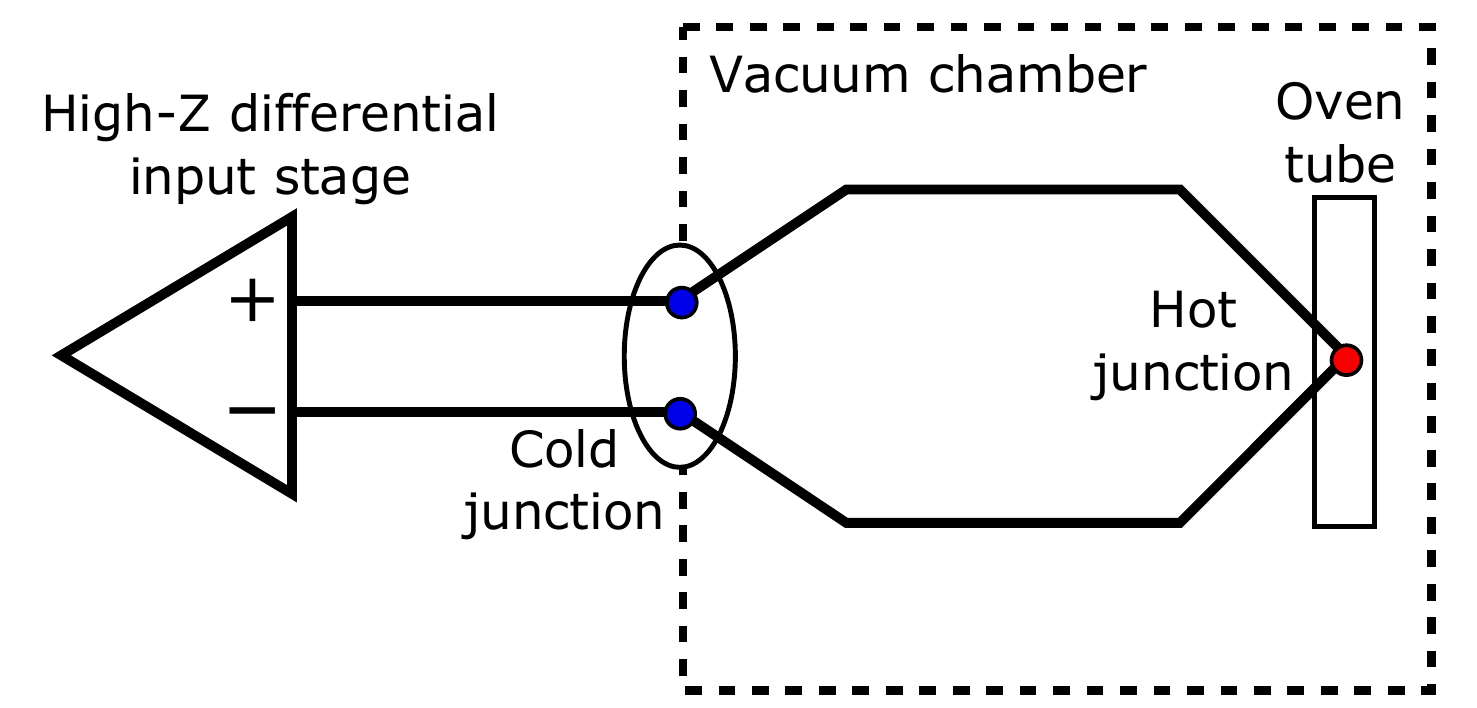}
\caption{Thermocouple circuit. The thermocouple hot junction is spot welded close to the oven aperture, while an effective cold junction is formed where the thermocouple wires are attached to a pair of copper feedthrough wires on the vacuum chamber. The voltage generated by the thermocouple is measured with a high-impedance differential input stage, ensuring no oven current flows to ground via the thermocouple wires.}
\label{fig:circuit}
\end{figure}

In order to provide a measure of the temperature of the oven, a K-type thermocouple junction is spot-welded to the tube wall close to the aperture. The thermocouple wires are connected to two terminals of a standard high-voltage ceramic and copper feedthrough, which forms the cold junction of the thermoelectric circuit, as shown in Figure~\ref{fig:circuit}. Electrical connections for the heater current to a second pair of feedthrough terminals are made via two copper wires held with screw contacts to the flexure clamps.
This construction, shown in Figure \ref{fig:oven_assy}, is made from easily machined parts yet ensures the atomic oven can be accurately aligned to the trap, improving reproducibility.
Furthermore, the low thermal mass of the tube permits the oven to be rapidly heated to its operating temperature, while its low thermal conductance prevents excessive heating of the surrounding structures.
A variation on this design consists of two similar oven tubes held by individual pairs of flexure clamps, in turn mounted to a single, wider Macor support. This simple extension allows independent loading of two ion species from one compact assembly. The measurements we present here were made with the single-oven design, but are representative of the behaviour of either variation.

\section{Control}

The temperature of the oven is controlled with a digital feedback loop implemented in a microcontroller. The voltage across the thermocouple junction is measured via a high-impedance differential analogue input stage, sampled with an analogue-to-digital converter at a rate of \SI{1}{kHz}, and the current is supplied by a buck-type voltage regulator controlled with a pulse-width modulated signal from the microcontroller. The resistance of the oven tube when cold is approximately \SI{200}{\milli\ohm}, with additional lead resistances totalling approximately \SI{50}{\milli\ohm}. The controller circuit can supply up to \SI{20}{A} into this load, while approximately \SI{3}{A} is required to maintain the tube at a temperature of \SI{400}{\celsius}. With the feedback loop optimized for a fast rise-time with no over-shoot, it is possible to heat the oven to \SI{400}{\celsius} in $\approx\SI{3}{s}$.

Since the vapor pressure is an exponential function of temperature, the absence of over-shoot in the step response is crucial to ensure that the atomic flux does not exceed the limits for safe operation. This is important because one of the design goals is to minimise deposition on the trap structure from the atomic beam.

\begin{figure}
\includegraphics[trim={2cm 17cm 3cm 1cm},clip,width=\columnwidth]{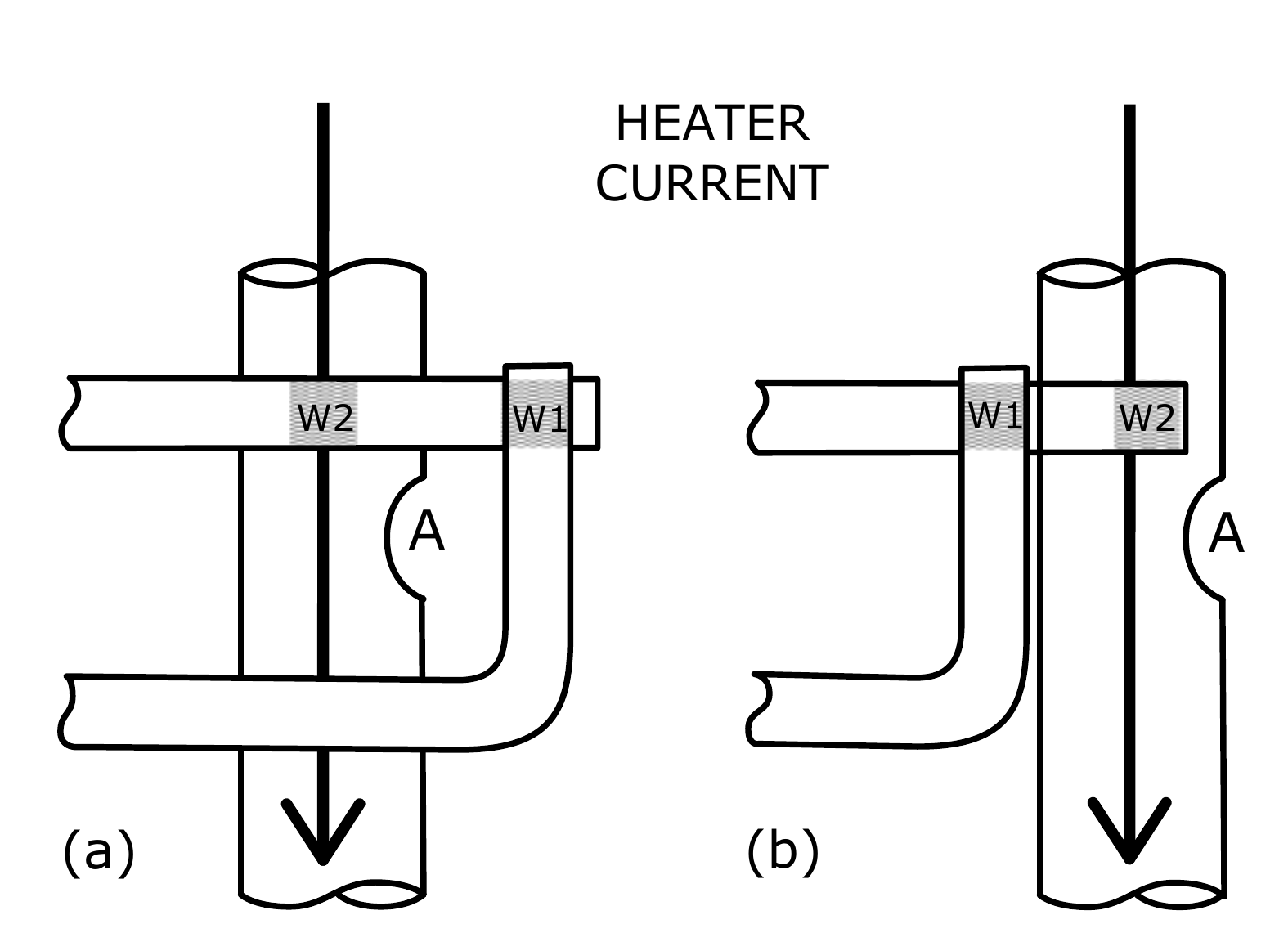}
\caption{Details of thermocouple weld. The thermocouple circuit is formed by two wires spot welded at junction W1. (a) If the spot-weld to the oven tube (W2) is located within the thermocouple circuit, ohmic voltages generated by the fraction of oven current flowing through the weld are summed with the thermoelectric voltage, introducing a current-dependent measurement offset. (b) If the junction W2 is located outside the thermocouple circuit, no significant oven current flows through this circuit and no additional offset voltage is seen. The thermocouple junction must nevertheless be located close to the oven aperture (A) to minimise the response time of the sensor.}
\label{fig:weld}
\end{figure}

If the thermocouple is spot-welded directly to the tube, we observe that the temperature measurement suffers from a systematic error which is proportional to the current flowing through the tube.
The coefficient of proportionality is strongly dependent on the geometry of the spot-weld and is minimized when the thermocouple wire is welded perpendicular to the direction of current.
This systematic error is understood to be due to oven current flowing not only through the walls of the tube, but also through the thermocouple wire material proximal to the spot-weld. Any current flowing through this weld region with direction parallel to the axis of the thermocouple wire will introduce an additional ohmic voltage indistinguishable from the thermoelectric effect.
The effect may be avoided by separating the thermocouple junction and oven tube welds, and ensuring the oven tube weld is located outside the thermocouple circuit, as shown in Figure~\ref{fig:weld}.
Alternatively, by modulating the current faster than the thermal response time of the oven, the current-to-voltage coefficient can be determined. With the coefficient known, the control system can compensate for this systematic error by feeding forward on the drive current.

\section{Measurement of atomic flux}

\begin{figure}
    \center
    \includegraphics[trim={5cm 7cm 7cm 3cm},clip,width=\columnwidth]{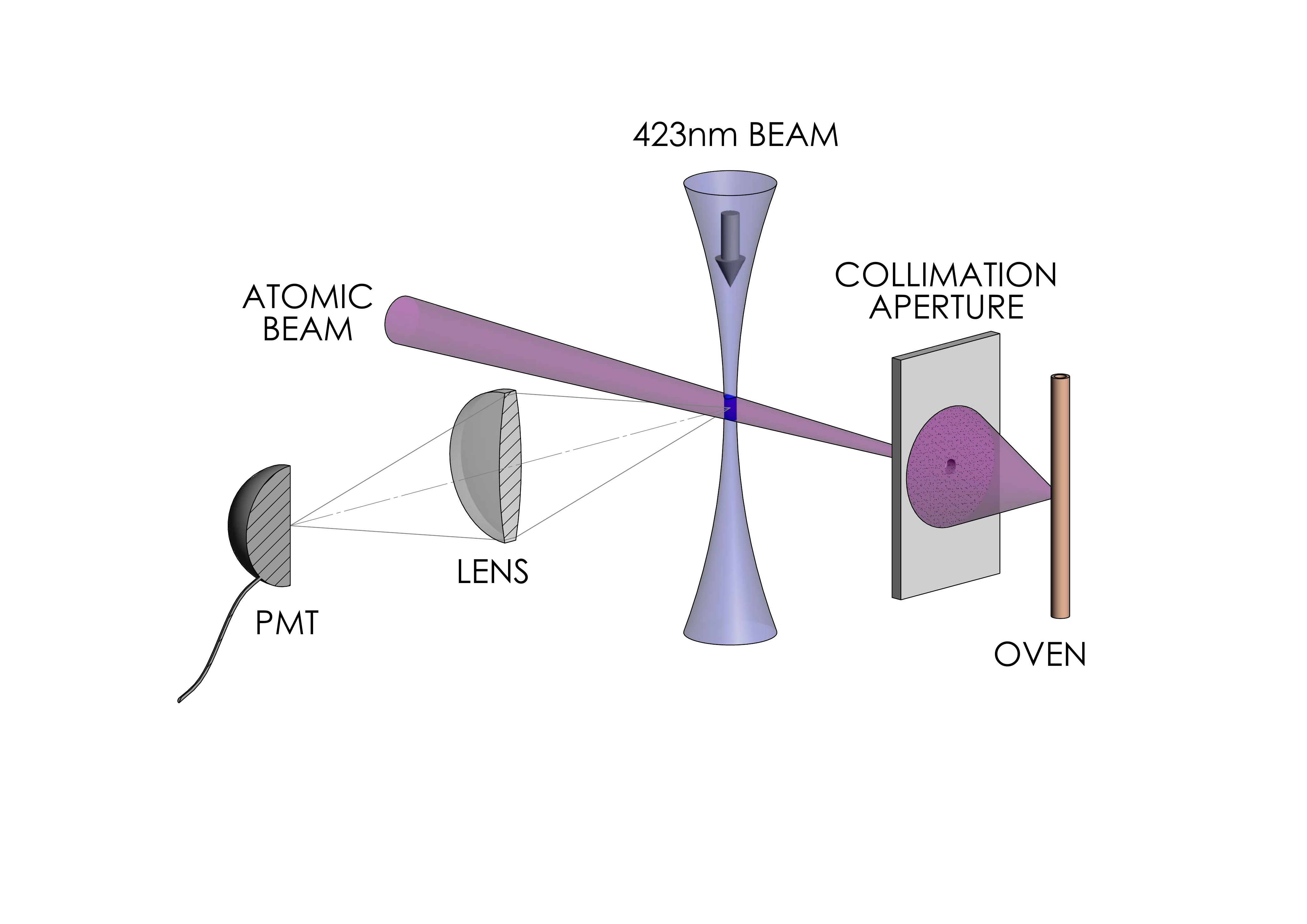}
    \caption{Schematic of the fluorescence measurement setup for determination of atomic flux. The excitation laser is orthogonal to the atomic beam to minimize Doppler effects. The atomic fluorescence is imaged onto a photomultiplier (PMT) by an objective lens.}
    \label{fig:exp_layout}
\end{figure}

To validate our design as an atomic source, we measure the atomic flux by laser resonance fluorescence as a function of time after oven turn-on and turn-off. The oven is placed inside a vacuum chamber at approximately \SI{E-6}{mbar}. A laser at \SI{423}{nm} is focused to a waist of \SI{100}{\micro\metre} at a position approximately \SI{10}{mm} from the oven collimation aperture. The angle of the laser beam is aligned so that it is perpendicular to the direction of the atomic beam in order to minimize Doppler broadening and shifts of the atomic resonance with respect to the laser, as shown in Figure \ref{fig:exp_layout}.

\begin{figure}
    \center
    \includegraphics[scale=1]{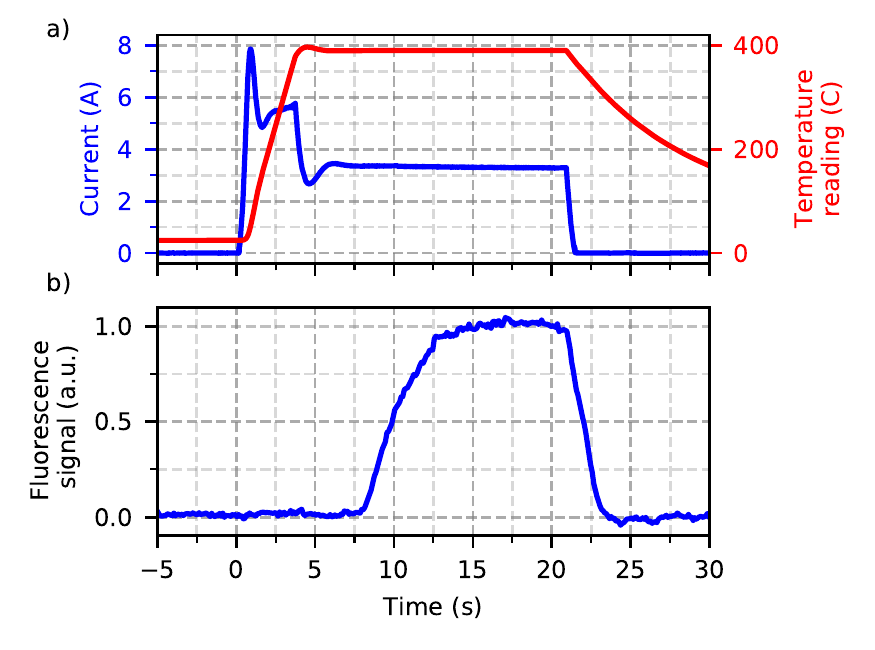}
    \caption{Performance of oven controller: (a) Oven current (blue) and thermocouple temperature (red) vs. time. (b) $^{40}\textrm{Ca}$ fluorescence signal vs. time. Note the delay between the thermocouple temperature reaching steady state and the onset of the emission.}
    \label{fig:graph}
\end{figure}

In Figure~\ref{fig:graph} we show the current, temperature and detected fluorescence when the oven is turned on for \SI{21}{\second}. The controller regulates the current to reach a steady temperature at the thermocouple in \SI{4}{\second} with minimal overshoot. Detected fluorescence shows that the emission of atomic flux begins after \SI{8}{\second}, increasing to its equilibrium rate after \SI{12.5}{\second}. We attribute the delay in this response to the thermal mass of the calcium and the relatively poor thermal conductivity between the calcium and the walls of the oven. When the current is switched off, the atomic flux drops to zero after \SI{2}{\second}. Note that for these experiments the optical detection efficiency has not been precisely calibrated and the exact number density of the atomic beam is not known, but was made substantially greater than would be necessary for loading ions in order to generate a high rate of fluorescence. The emission rate is can be reduced to any value desired via adjustment of the temperature setpoint.

The loading rate in a real trap will depend strongly on the trap depth and details of the photoionisation process, which both vary greatly between experiments. As such we have not conducted a comprehensive study of the performance of this oven in a particular trap, which would be less useful than the measurements we present of the stability and temporal response of the flux produced. However, we have used this oven design to load a surface trap in our latest experiments, and observe that the average loading time is similar to the equilibration time of the atomic flux.

\begin{figure}
    \center
    \includegraphics[width=\columnwidth]{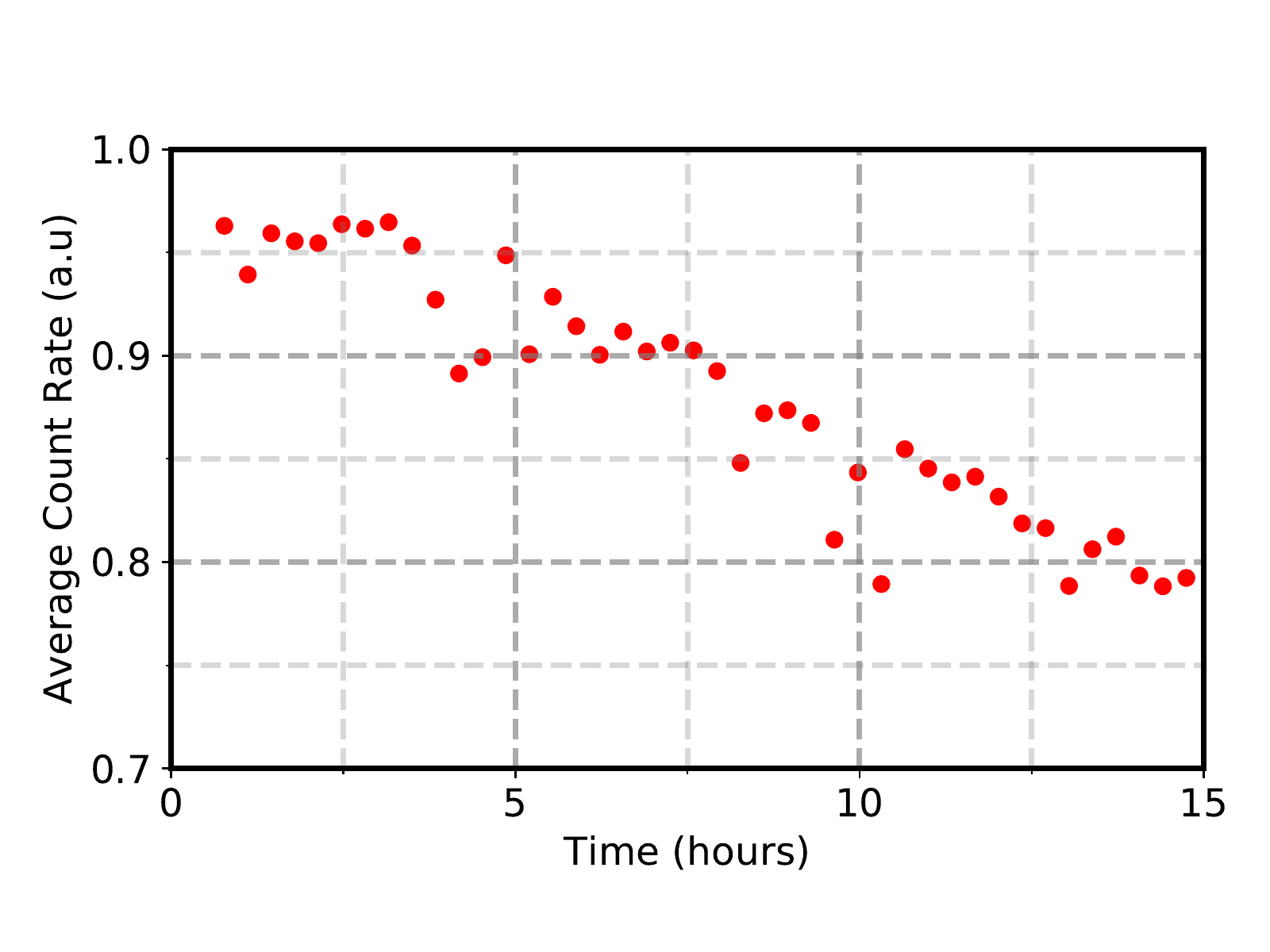}
    \caption{Average fluorescence count rate of repeated firing cycles over a cumulative on-time of 15 hours. The gradual reduction in count rate is attributed to depletion of the calcium in the oven, and corresponds to an integrated flux sufficient to load $\sim10^7$ ions in a typical trap. The fast noise is due to laser drift in between measurement cycles.}
    \label{fig:lifetime}
\end{figure}

Repeated firings of the oven at a given temperature set point yield stable and consistent results over a long period of time. In order to test the long term stability of the atomic flux generated by the oven, we fired the oven repeatedly at close to 100\% duty-cycle at over a period of 15 hours. The steady-state fluorescence count rate was measured at regular intervals, as shown in Figure~\ref{fig:lifetime}. The atomic flux appears stable for the first few hours, before starting to steadily decrease with a time constant of approximately 60 hours. We note that after 15 hours, the total number of atoms which have passed through the central $(\SI{20}{\micro\metre})^3$ region of the beam is estimated to be $\sim10^{10}$. In our surface traps this would be sufficient to load a total of $\sim10^7$ ions, many more than are necessary over the lifetime of any typical experiment. Considering the geometry of the collimator, this integrated flux corresponds to emission of a significant fraction of the $\SI{2}{\milli\gram}$ of calcium loaded into the oven, and the reduction in flux observed is likely due to preferential depletion of the calcium from hotter areas of the oven walls.

\begin{figure}
    \center
    \includegraphics[scale=1]{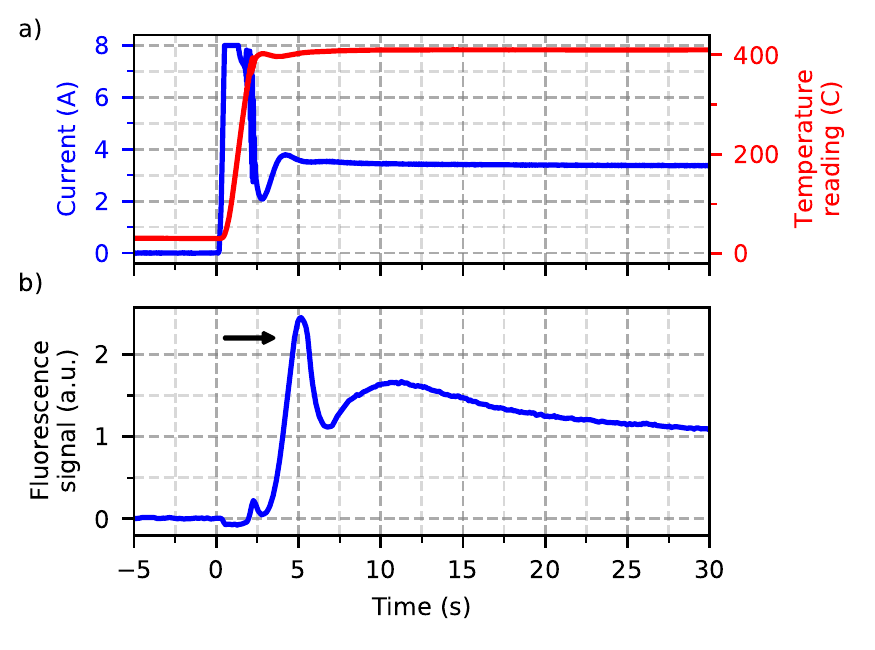}
    \caption{Behaviour of oven after heat treatment. (a) Oven current (blue) and thermocouple temperature (red) vs. time. (b) $^{40}\textrm{Ca}$ fluorescence signal vs. time. The initial strong peak (arrowed) is indicative of redistribution of calcium within the oven tube, which could allow for a reduction in the turn-on time of the oven.}
    \label{fig:heat_treatment}
\end{figure}

If, prior to testing, the oven is heated to a somewhat higher temperature, a distinct difference is seen in the emission dynamics in subsequent tests. Figure~\ref{fig:heat_treatment} shows the emitted flux in a test similar to that shown in Figure~\ref{fig:graph}, but conducted after first heating the oven to \SI{450}{\celsius} and allowing it to cool. Here, two distinct processes are observed over different timescales. The onset of emission occurs after just \SI{2}{\second} and rapidly reaches a flux more than twice as high as the equilibrium level from the previous test. This initial `burst' of atomic flux tracks the oven temperature much more closely with minimal delay; the sharp peak can be correlated with the small overshoot of the oven temperature during turn-on. As the oven temperature stabilizes, the flux remains high but decays towards its equilibrium level over \SI{20}{\second}. We hypothesize that pre-heating the oven to the higher temperature leads to the deposition of a thin layer of calcium over the inside of the oven cavity. This layer has a low thermal mass and is in direct contact with the oven walls, tracking their temperature very closely and causing the increased initial rate of emission. As the oven remains on, the main calcium mass heats and begins to vaporize, while the deposits on the walls are rapidly depleted. Subsequent repeats of the experiment exhibit the normal behaviour, as shown in Figure~\ref{fig:graph}, while repeating the high temperature pre-treatment leads again to the dynamics we have just described.

This behaviour suggests a means to produce an oven with a faster response. If the calcium inside were heated under vacuum to melting point (\SI{842}{\celsius}) before drilling the oven aperture, a thicker but relatively uniform and well thermally-contacted layer could be deposited on the inside of the oven tube. For the experiments described here, this would allow for the production of controlled atomic flux in \SI{2}{\second}.\\

\section{\label{conclusion}{Conclusion}}

We have presented and tested a design for an atomic oven suitable for loading of ion traps, which is compact and easy to reproduce. By controlling the oven current to stabilize the temperature of a thermocouple mounted next to the oven aperture with a digital feedback loop, we are able to reach a steady level of atomic flux in \SI{12}{\second} with no overshoot. We have successfully used this oven system to load ions in a surface trap.

The closed-loop control allows the the oven to be run at a higher temperature for a shorter period of time when loading ions.
Because the number density in the atomic beam is a strong exponential function of temperature, the time-integrated flux required to load a single ion can thus be achieved with less total energy delivered to the trap environment. Combined with the low thermal mass and conductivity of the oven, this will reduce heating of the neighbouring trap and vacuum systems, reducing thermally-driven electrical and mechanical variations.

We hope other researchers may find this design useful for their own work; detailed schematics are available on our online repository\cite{GrimRepo}.\\

\section*{Acknowledgments}
This work was supported by the EPSRC Quantum Technology Hub in Networked Quantum Information Processing. The authors thank Andrew Steane for helpful discussions and input. TGB is employed by Cold Quanta UK Ltd.\\


%

\end{document}